\documentclass[fleqn,10pt]{wlscirep}
\usepackage[utf8]{inputenc}
\usepackage[T1]{fontenc}

\title{Virtual reality for 3D histology: multi-scale visualization of organs with interactive feature exploration}

\author[1]{Kaisa Liimatainen}
\author[2]{Leena Latonen}
\author[1]{Masi Valkonen}
\author[1] {Kimmo Kartasalo}
\author[1,3,*]{Pekka Ruusuvuori}
\affil[1]{Tampere University, Faculty of Medicine and Health Technology, Tampere, FI-33014, Finland}
\affil[2]{University of Eastern Finland, Institute of Biomedicine, Kuopio, FIN-70211, Finland}
\affil[3]{University of Turku, Institute of Biomedicine, Turku, FI-20014, Finland}

\affil[*]{pekka.ruusuvuori@utu.fi}

\begin{abstract}
Virtual reality (VR) enables data visualization in an immersive and engaging manner, and it can be used for creating ways to explore scientific data.
Here, we use VR for visualization of 3D histology data, creating a novel interface for digital pathology. Our contribution includes 3D modeling of a whole organ and embedded objects of interest, fusing the models with associated quantitative features and full resolution serial section patches, and implementing the virtual reality application.
Our VR application is multi-scale in nature, covering two object levels representing different ranges of detail, namely organ level and sub-organ level.
In addition, the application includes several data layers, including the measured histology image layer and multiple representations of quantitative features computed from the histology.
In this interactive VR application, the user can set visualization properties, select different samples and features, and interact with various objects. In this work, we used whole mouse prostates (organ level) with prostate cancer tumors (sub-organ objects of interest) as example cases, and included quantitative histological features relevant for tumor biology in the VR model. Due to automated processing of the histology data, our application can be easily adopted to visualize other organs and pathologies from various origins. Our application enables a novel way for exploration of high-resolution, multidimensional data for biomedical research purposes, and can also be used in teaching and researcher training.

\end{abstract}
\begin{document}

\flushbottom
\maketitle
%
%
\thispagestyle{empty}


\section*{Introduction}



Histological assessments remain the cornerstone of pathological diagnostics and research. Currently, traditional microscopy is being increasingly substituted by digital pathology, in which tissue sections are imaged to digital high-resolution whole slide images (WSI). In addition to enabling detailed analysis of the histological samples, the WSIs enable digital analysis and quantification of histological features. Although still nontrivial, the digitalization also allows alignment of serial sections into a common coordinate space, resulting in histology data represented in 3D and thus replicating the three-dimensional nature of the original tissue \cite{roberts2012toward,magee2015histopathology,kartasalo2018}. Advanced visualization methods that reach beyond observing single images on a 2D screen are required to obtain full benefits from these three-dimensional datasets. 

In addition to modeling the tissue based on histological sections only, also fusion models of histological serial sections and native 3D imaging modalities such as $\mu$CT \cite{dullin2017muct} or MRI \cite{gibson20133d,schmidt2004volume,bart2005mri} are possible. Even manually created 3D models representing epithelium with textures acquired from histological samples have been made for teaching purposes \cite{sieben2017histology}. 
However, applicability of 3D imaging modalities depends on the physical sample size and such measurement devices are not always available, while manual modeling from histology data is time-consuming and subjective. Further, unlike 3D imaging, modeling based on serial sections is possible also for already sectioned and processed samples, e.g. archived clinical sample sets. 

One challenge in 3D histology visualization is the large volume of the data - the size of full resolution images is typically in the range of gigapixels. Reading multiple full resolution WSIs to a computer memory is time-consuming and makes data visualization slow due to long loading times. Modeling of 3D histology data creates a further challenge, not to mention adding feature representations on top.  


Virtual reality (VR) can be used for realistic and immersive data visualization in various applications of medical imaging. 
However, to the best of our knowledge, a tool for modeling and visualization of sparse 3D histology data in VR has not yet emerged. 
Common 3D visualization tools in e.g. Python or MATLAB can be used to visualize the data and quantitative features, but they do not provide the possibility to fully interact with the data, and also do not allow detailed exploration. Furthermore, trivial use of histological visualization requires user-friendly interfaces accessible to pathologists and biomedical researchers without computational knowledge.
With well-thought design choices, VR can provide a comfortable and compelling experience for viewing and interacting with scientific data. It is emerging as a popular tool for training simulations in medicine  \cite{mcguire2018competency,liebig2018metric} and for surgical planning \cite{shirk2019effect,frajhof2018virtual}.
In bioinformatics, virtual reality has been used for e.g. visualization of complex molecular structures \cite{wiebrands2018molecular,chavent2011gpu,trellet2018semantics,biere2018heuristic} and tracing neurons \cite{usher2017virtual}.
3D representations in VR can concretize complex structures in a manner that is not possible to achieve with a 2D screen - the same holds true for 3D printing \cite{liimatainen20193d}, but VR enables more flexible and detailed visualizations. The spatial connectivity of objects of interest is easy to grasp in VR. In addition, datapoints from e.g. quantitative features can be embedded to the 3D representation for visualization and exploration of additional information.

In this study, we use sparse stacks of histological sections that span through whole organ samples to create realistic 3D organ models in virtual reality. 
We combine the created whole organ models with models of internal sub-organ objects of interest, and embed quantitative feature visualizations to the models. In our virtual reality application, the user can interact with features to see image patches from full-resolution serial sections they are connected to, and further, study the objects of interest in corresponding sub-organ level with embedded serial sections.

Important aspects that have been taken into consideration are performance optimization for the VR application, automation in data processing and level construction, and comfortable user experience. 
In our prototype application, we use objects in two different scales, with mouse prostates as whole organs and prostate cancer tumors as objects of interest (sub-organ), associated with quantitative features studied for their relevance for the organ anatomy and tumor biology. With our use-case, we show how the organ level view in the VR application enables detection of cancer growth pattern differences, validated in the detailed histology level view, and explored for the associated quantitative details. Our application presents as a novel interface for digital pathology, and it can be used for education and teaching as well as for research purposes.

\section*{Application and results}

\begin{figure}
    \centering
    \includegraphics[width=\linewidth]{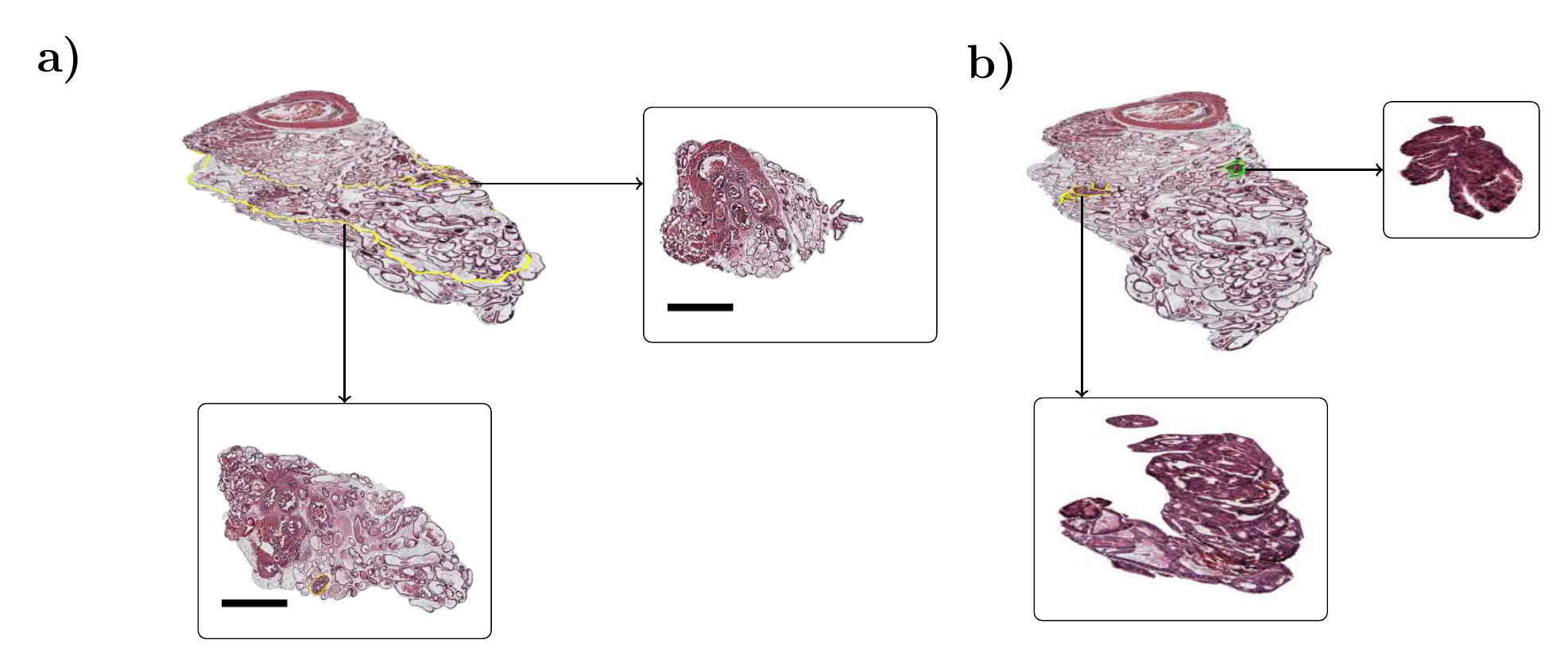}
    \caption{Image stack composed of histological sections from a mouse prostate visualized in Blender. Two example sections are shown (scale bar = 1mm) (a). Manually annotated tumor locations with two tumor image stacks visualized (b). }
    \label{fig:imagestacks}
\end{figure}

To visualize a whole organ from serial sections, we used sparsely spaced histological WSIs representing whole tissues. 
Histological image stacks, rendered with Blender \cite{blender}, are visualized in Figure~\ref{fig:imagestacks}. In Figure~\ref{fig:imagestacks} (a) example sections from the stack, outlined in yellow, are presented. Manually annotated tumors are presented in Figure~\ref{fig:imagestacks} (b), outlined in the image stack and rendered individually. 

The application prototype consists of two different levels. At the organ level, the user can view the whole organ with its structures (Fig.~\ref{fig:application} (a)), pathologies, and even the associated quantitative features (Fig.~\ref{fig:application} (b-c)), as well as examine the features interactively (Fig.~\ref{fig:application} (d)). In our use-case, this is the level where the prostate with tumors and the computationally calculated features can be viewed and interacted with. At the sub-organ level, the user can interact with single areas of interest (in the example case the tumors; Fig.~\ref{fig:application} (e)) and view the serial sections and features computed from these interest locations (Fig.~\ref{fig:application} (f)). Application properties and navigation in both levels is demonstrated in Supplementary Video S1 \footnote{https://github.com/BioimageInformaticsTampere/The-Virtual-Prostate}.

Patches of full resolution serial sections are visualized in both levels. Since loading of entire full resolution serial sections is time-consuming, the sections are cropped to smaller patches representing objects of interest to enable image loading without delay. To create a comfortable user experience, controllers are fully utilized for easy navigation and data exploration. User friendly menus are added to levels for full control on selecting the data and visualization method. 
In addition, ambient music was created for both levels to make the application even more immersive. 
New samples can be added to the application without manual effort due to automated data processing and dynamic loading of data to levels.

\subsection*{Organ level view: Virtualization of a whole prostate model}

In the main level of the VR application, prostate outline is visualized together with tumors (Fig.~\ref{fig:application} (a)) and quantitative features (Fig.~\ref{fig:application} (b-c)), as shown in Supplementary Video S1.
The interactive spatial visualization of the whole organ allows intuitive examination of the tumor locations within the organ as well as tumor shapes and growth patterns, in a manner not possible through conventional 2D approaches. Furthermore, through visualization of different quantitative feature values spatially, it is possible to examine which features are connected to and how to the areas of interest (in this case the tumors and their growth patterns) and other parts of the organ. Here, the tumors are rendered as transparent for clear visibility of the structures and features inside the tumor. 
The features can be selected from a drop-down menu and a slider is used to define the threshold, which, in order to effectively handle outlier values, is defined as percentiles instead of absolute feature values. Only the features belonging to percentile with highest values are visible.
Feature visibility changes instantly when slider value is adjusted, making it easy to find a suitable threshold. 
With easy feature selection and instant thresholding, this level helps in glancing data volumes in exploratory manner and to define visually which features can be used to describe specific properties of the tumors or the prostate. Especially for an expert biologist, the image patches corresponding to specific feature spheres can provide insight to which features are connected to specific tissue structures or histological appearances. 

There are two different modes for feature visualization: visualizing all features simultaneously as particle systems (Fig.~\ref{fig:application}(b)), and visualizing features whose values are above or below a given threshold as spheres (Fig.~\ref{fig:application}(c)). The color of a sphere and particles defines the strength of the feature, based on a user-selected colormap. With the particle system, all features can be visualized at once, since small particles are light to render and do not block the view as solid objects would. It is also possible to plot all spheres at once, however, with large volumes this may cause lag and make the level too crowded for interaction with spheres, possibly decreasing quality of user experience.  
To examine a feature in a given area, the user can grab a feature sphere to expose the feature value and visualize an image showing the patch from which a given feature was computed from as a cropped area from full resolution histological sections. The user can also zoom the image with two-handed gestures, similarly to two-finger zoom on touch displays. To make particle systems more informative, the feature value affects particle's spawn rate and lifetime. When a feature's value is high, particles spawn more frequently and stay alive longer, making these areas more occupied by particles. Thus, areas with higher feature values can be distinguished even from a distance, as demonstrated in Figure~\ref{fig:application}(b) and Supplementary Video S1.   

\begin{figure}
    \centering
    \includegraphics[width=\linewidth]{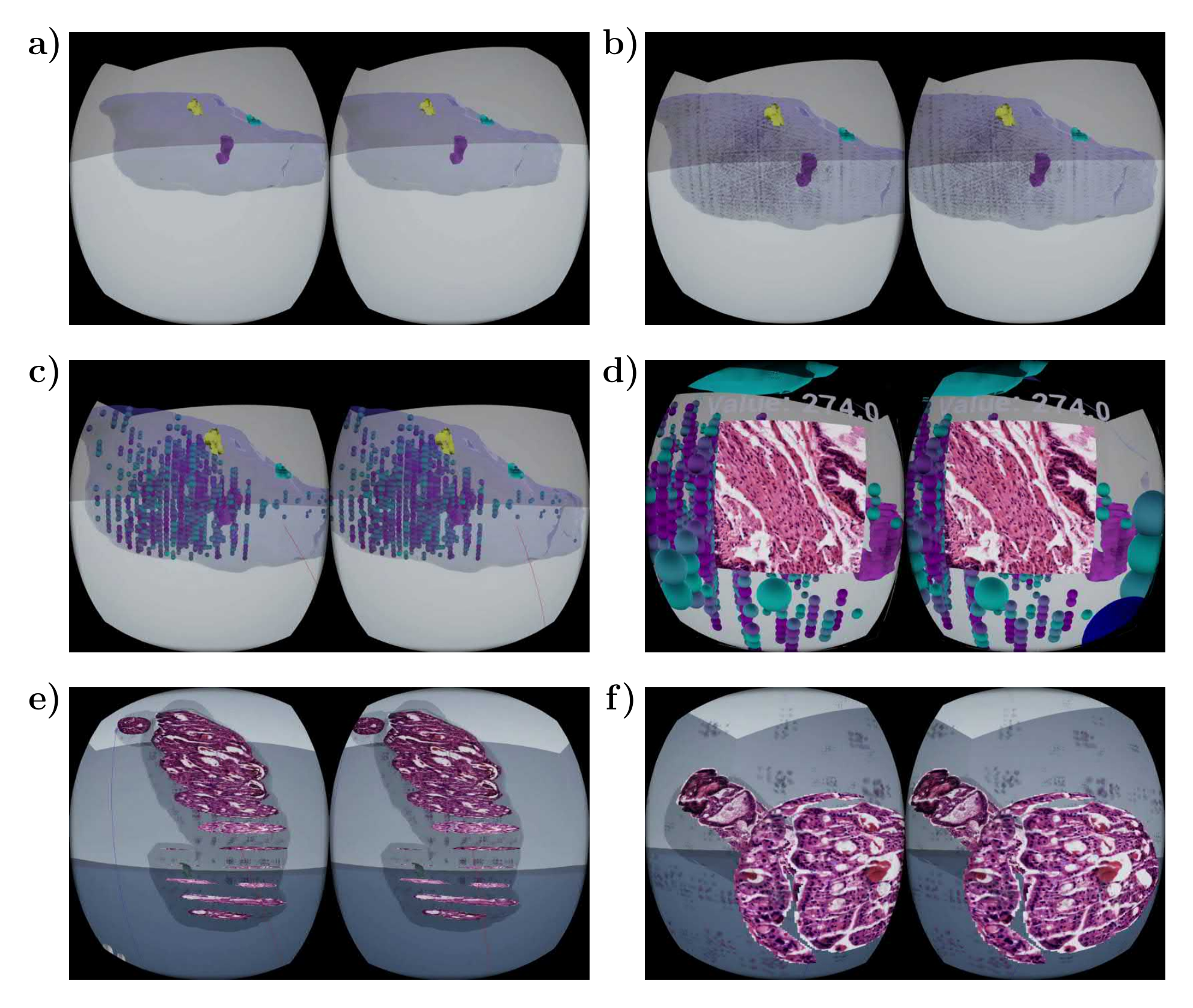}
    \caption{VR application view examples were captured during gameplay. Prostate and tumor models are visualized in organ level view (a). Features can be presented as particle systems (b) or grabbable spheres (c). In organ level view, the user can interact with tumors (left hand) and features (right hand). When a feature sphere is grabbed, the corresponding image patch is visualized (d). A tumor with full-resolution H\&E stained serial sections and with associated quantitative features is visualized in tumor view  (e). The tumor can be rotated and serial sections made invisible when studying the tumor in detail (f). }
    \label{fig:application}
\end{figure}


\subsection*{Detailed view on sub-organ region of interest: Multiscale virtual tumor view}

In the main (organ) level, the user can select an object of interest which is visualized and can be examined in more detail at a specific sub-organ level. In this level, the user can study how the tumor appears in the histological sections, and which features are associated to that tumor or its parts (Fig.~\ref{fig:results_tumor}). Inside the tumor, each histological section of the tumor is visualized at their correct 3D spatial location. This allows, for example,  exploration of the tumor histology according to the spatial location within the tumor, aiding in evaluation of tumor heterogeneity and research on associated tumor growth patterns.

The user can adjust the rotation and vertical position of the tumor directly by holding designated buttons and rotating or moving the controller. This design enables easy adjustment of the tumor model to get the best, eye-level view of the histological sections inside.
Histological sections can then be inspected in detail. To enable browsing sections back and forth, a section can be made invisible such that the consecutive section is in full view, enabling studying all the sections in a tumor.


The histological images were acquired from full resolution sections, and with a high-resolution VR headset they can be studied in high detail. The tumor scale can also be changed, which simultaneously changes the scale of serial sections. 
Similarly to the organ level, features can be plotted inside the tumor as particles and as transparent spheres. If the user selects to view the features, the feature spheres are rendered transparent to enable visibility of the histological sections  underneath. This allows easy visualization and examination of which features are associated with certain histologies.


\subsection*{Volumetric visualization and tumor exploration in prostate cancer: Tumor growth pattern differences associated to spatial features beyond histology}

The usability and utilization of our VR model was tested with prostate tissues to study the spatial organization of the tissue as well as tumors of prostate cancer. Prostates of mice genetically heterogeneous with Pten tumor suppressor (\emph{Pten+/-}; \cite{cristofano1998pten}) form intraglandular, noninvasive high grade prostatic intraepithelial neoplasia (PIN) tumors which were here studied at the age of 11 months. To evaluate heterogeneity in morphology of the tissue and location of the tumors between different prostates, six prostate samples were compared by visualizing a 3D model of the whole organs. In Figure 3, three of these visualizations are presented as still images from three different angles each. The VR application made exploration of the tissue shape and locations of areas of interest simple and intuitive.

Each prostate had several tumors, and their amount, location and size were found to vary greatly between the samples – a variation which is not evident from 2D histology especially when routinely only a few sections per mouse prostate are examined in such cases. These tumors are currently also too small to be visualized by live imaging methods such as PET, making this approach unique in visualizing them as 3-dimensional structures. Most tumors in these samples localize to the lateral lobes of the prostate, which is previously known based on 2D histology and also evident here in relation to the anatomical shape and orientation of the tissue. Each prostate sample had one tumor that was clearly larger than the rest. Tumors were found to often localize near the outer borders of the prostate. Especially, the largest tumors were noted to reach to the outer areas of the tissue. Since mouse prostate is an unencapsulated organ with edges free to expand to the abdominal cavity, the localization effect is likely due to growth pressure in the tissue being smaller towards the edges, thus giving a growth advantage.

The specific sub-organ view was used to study each tumor in more detail. This mode allows easy comparison between tumors to examine tumor heterogeneity and the associated features. In Figure 4, six tumors with various shapes and sizes are presented as examples, showing views of the full-resolution histological sections from the tumor area together with their locations in the prostates. The histological sections visible inside the tumor at the correct locations, and the ability to select any given section for individual examination (see Fig 2f), enable detailed histological analysis of intratumoral appearance and heterogeneity. For example, areas of increased cellular density towards a tumor edge likely represent growth fronts, while the cellularly sparse and secretion-containing areas are more benign. From these visualizations it is easy to note how the tumors grow along the prostate glandular structures with occasional elongated and even bended shapes.

The prostate samples used here have previously been used to develop quantitative histological analysis to computationally study descriptive features and spatial heterogeneity in 2D histological sections \cite{valkonen2017analysis}. Here, we computed quantitative histological features and included in the VR to explore them in the 3D context. Visualization of the small image patches to explore the histological appearance that resulted in the quantitated values, enabled exploration of tissue heterogeneity both at the whole tissue and tumor levels. In Figure~\ref{fig:results_features}, still images for four features visualized in a prostate are illustrated as examples. By setting the threshold relatively high, the locations with the most intense feature values can be highlighted. While certain features have strongest signals at specific anatomical parts of the prostate, some are spread evenly throughout the prostate. While only a fraction of the computed features is intuitive in association of 2D histology, visualizing the features in VR enables better understanding of the heterogeneity of anatomical feature distributions and allows discovery of features linked to 3D anatomical properties of the tissue. As an example of an expected result, the amount of nuclei is highest anatomically near the urethra running through the prostate (visible as a distinct red circle on top of the image stacks in Fig.~\ref{fig:imagestacks}), representing a high density of cells in the urethral muscle layer, intraurethral glands and ducts, as compared to the sparse prostate glandular structures distal to the urethra (Fig.~\ref{fig:results_features} (a)). The exploration reveals several anatomically concentrated features not intuitively explained, such as energy (sum of squared elements) in the grayscale co-occurrence matrix (Fig.~\ref{fig:results_features} (b)). Certain non-intuitive features can clearly be observed to concentrate on certain anatomical locations and tissue types areas, such as histogram of oriented gradients features in Figure~\ref{fig:results_features} (c) and local binary patterns (Fig.~\ref{fig:results_features} (d), \cite{ojala2000gray}). While additional research is required to understand the potential biological relevance of many of these observations, the ability of these features to distinguish tissue areas and types is likely to become useful in future quantitative analysis of tissue alterations.




\begin{figure}
    \centering
    \includegraphics[width=\linewidth]{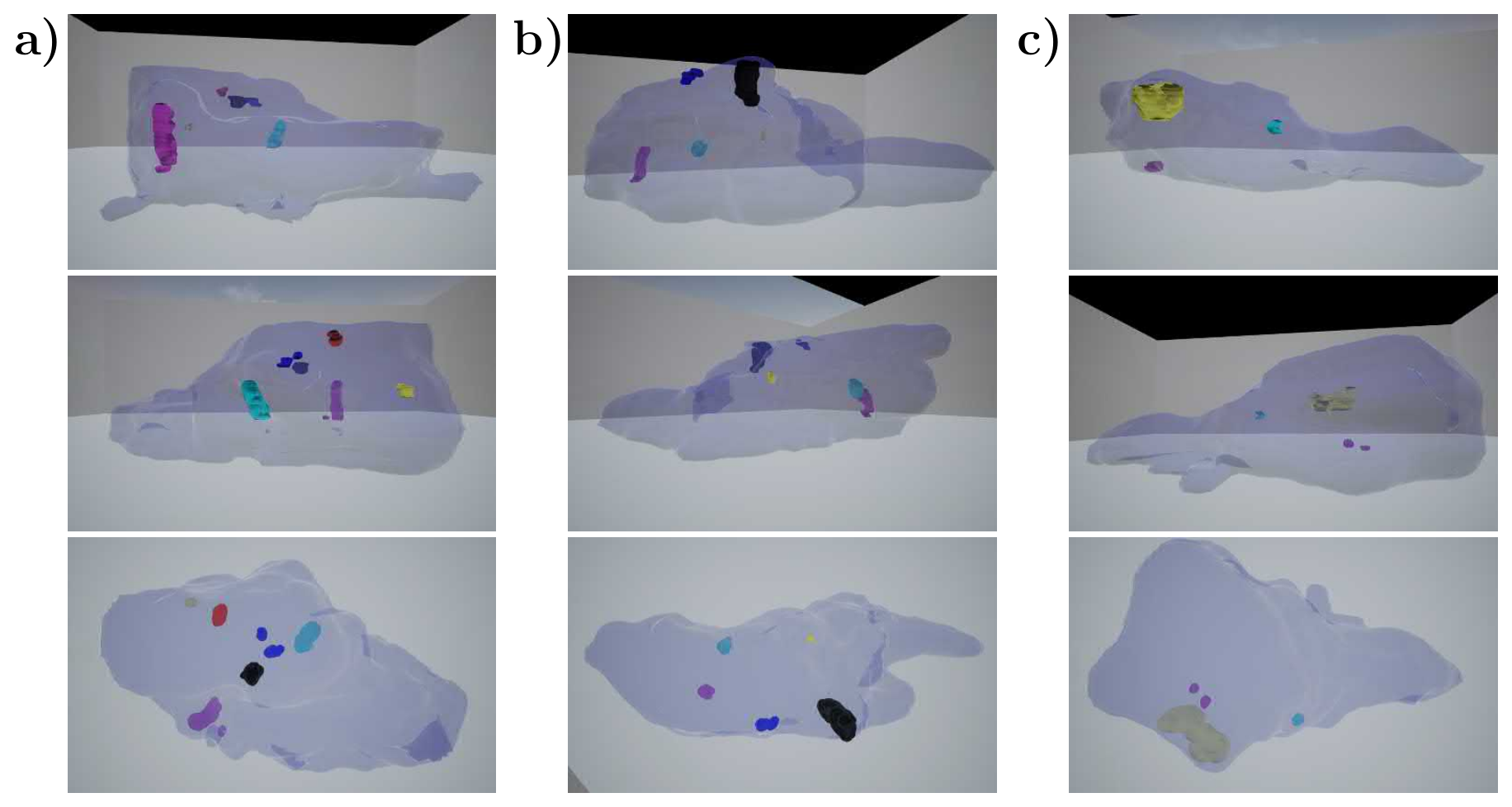}
    \caption{Three prostate samples with tumors visualized from different angles. Top and center figures are from both lateral sides of the prostates, bottom figures were captured from above, corresponding to the urethral side of the prostate. Height of each sample is approximately 1.5mm. Tumors have different colors for better separation. }
    \label{fig:prostates}
\end{figure}




\begin{figure}
    \centering
    \includegraphics[width=\linewidth]{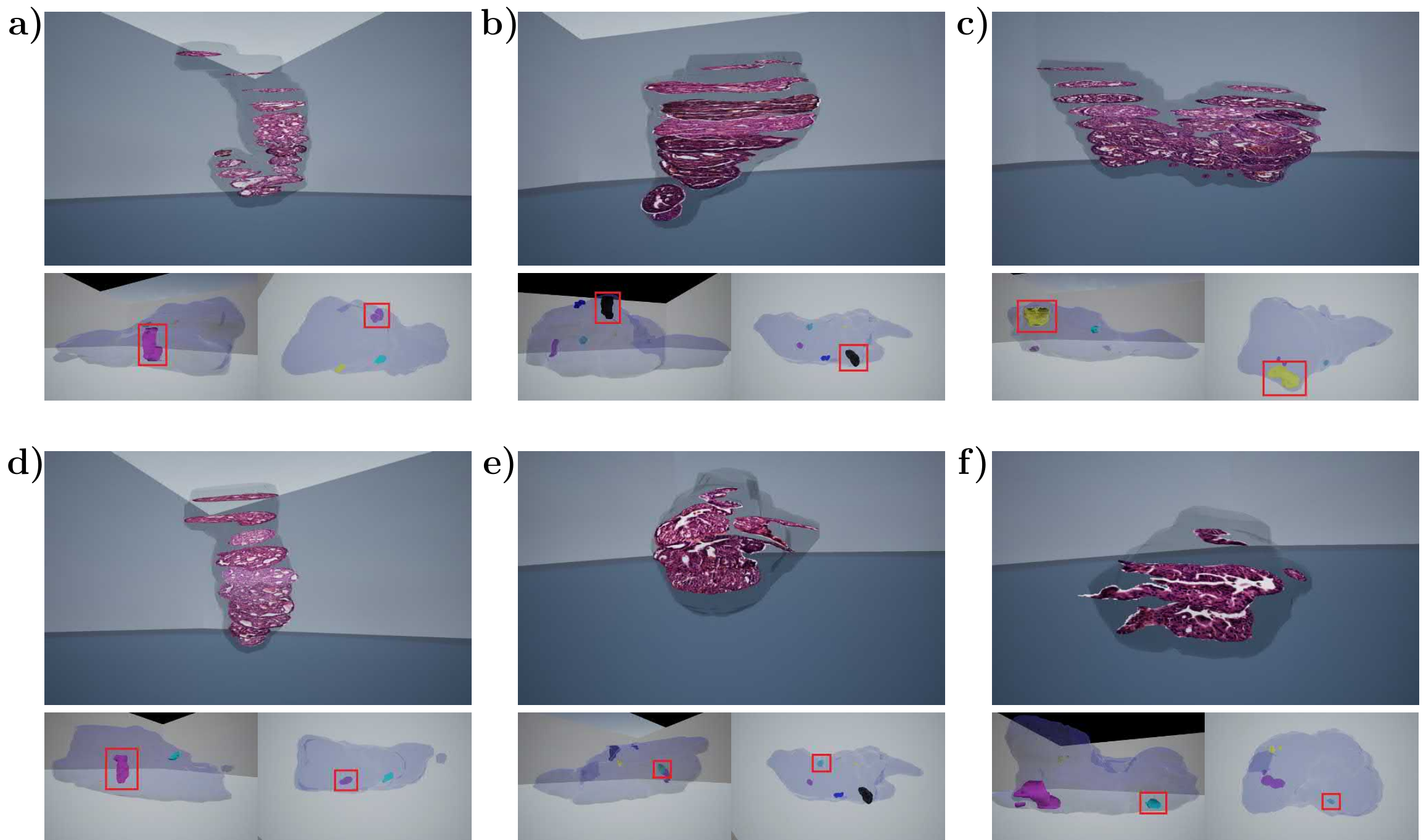}
    \caption{Tumors explored in detail in virtual tumor view. Examples of six tumors are shown, with the corresponding histological sections visible. Tumor locations in the prostate are presented below each tumor figure.}
    \label{fig:results_tumor}
\end{figure}




\begin{figure}
    \centering
    \includegraphics[width=\linewidth]{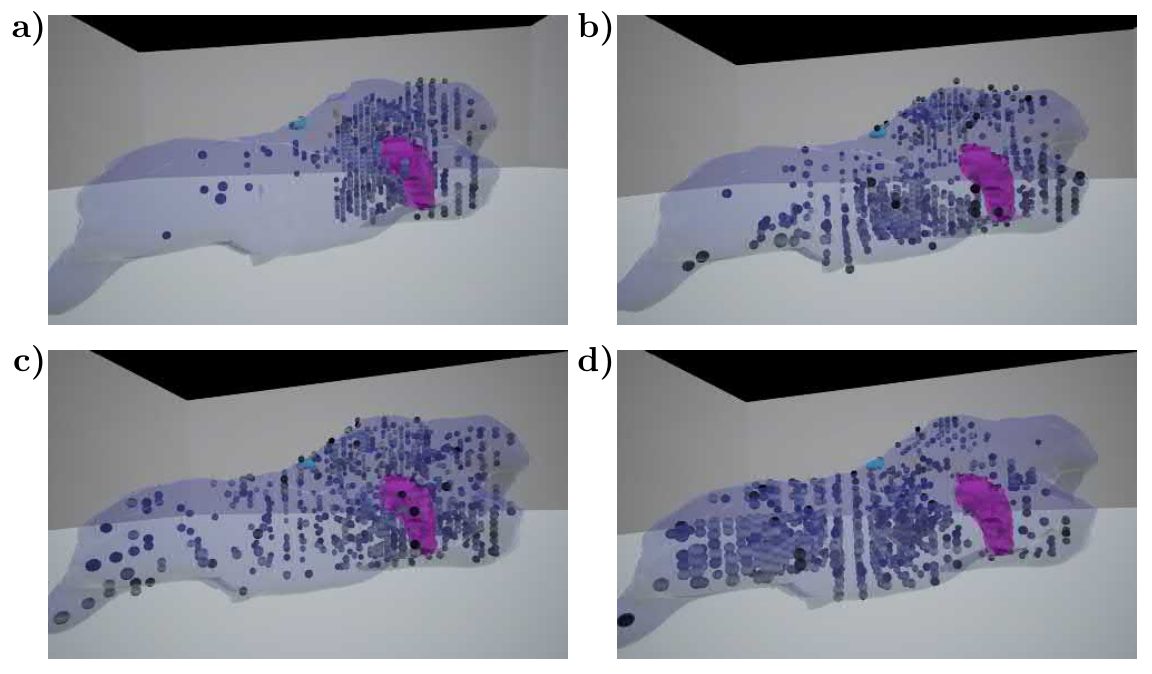}
    \caption{ Quantitative features presented in the 3D prostate volume demonstrate variation of signal intensities between features and in different anatomical parts of the prostate. Approximately 20\% of highest ranking features were plotted with bone colormap, where lighter colors represent higher values. Amount of nuclei is highest near urethra, representing high density of cells in ducts and urethral muscle layer compared to sparse prostate glandular structures distal to the urethra (a). Energy (sum of squared elements) in grayscale co-occurrence matrix is strongest at the top and bottom of the prostate (b). Histogram of oriented gradients features with 9 pixel window are spread around the prostate (c). Local binary patterns tend to have strong signals around lateral and ventral prostate (d). }
    \label{fig:results_features}
\end{figure}

\section*{Conclusion and discussion} 

In this research, we created a VR application for visualizing tissue histology in 3D. In addition to studying the gross anatomy of tissue and its regions of interest, our application can be used to study detailed histology of the samples. Furthermore, quantitative features computed from the serial sections were visualized. Our application is a novel tool for both research and teaching, with a fully immersive experience of walking inside an organ and interacting with the regions of interest and quantitative features to explore the tissue properties.

We used mouse prostates with prostate cancer tumors as a use-case. With the VR visualization, the tumors could be explored in ways that are not possible through standard 2D visualization of histological section – not even when several sections are utilized. The ability to observe the tissue and the tumors from multiple angles clearly enhances the information gained of how tumors are distributed in the tissue, their growth patterns and their heterogeneity. Considering that these tumors are too small to currently visualize by live imaging techniques, the observations enabled by the VR 3D visualization are previously unencountered and underline the usefulness of VR as a tissue exploration tool. Biologically, it is interesting that each prostate sample had one tumor that was clearly larger than the rest. The tumors in this \emph{Pten+/-} prostate cancer model are multifocal. Thus, observation of a “major tumor” within each prostate does not reflect a primary tumor from which the others have metastasized, but rather must represent a tumor that is either initiated early or has a stronger growth potential than the others. While these tumors were localized close to the edges of the tissue, they may have gained a spatial growth advantage due to growth pressure in the tissue being smaller where it is easier to expand towards the abdominal cavity. The observation, however, raises questions as to whether the largest, growth-advantaged tumor has a signaling capacity to keep the others’ growth slower in a similar fashion as has been indicated for primary tumors affecting the dormancy of its metastatic cells \cite{fares2020}.

Traditionally, quantitative features have been used in machine learning algorithms, and many methods enable defining which features have highest influence on classification results. We have previously used this prostate tumor material in a study where we analyzed spatial heterogeneity of neoplastic alterations in these prostate tumors in a 2D setting \cite{valkonen2017analysis}. We computationally quantified hundreds of features from the histological WSIs and used them to build a feature-based machine learning model to separate genetic alterations between tumors. Now, with our 3D VR application, we were able to easily explore firstly the spatial distribution of the important computational features to associate them to anatomical locations and, secondly, through visualization of the corresponding histology, study what properties they capture. 

In this study, the models were created based on sparse histological serial sections. Sparse section stacks have long distances between consecutive sections, which makes modeling challenging and results in models where discrete steps between sections are visible.
In addition to modeling prostates with tumors, we could create models of inner structures of the prostate, like glands and urethra. This would provide a more informative, comprehensive model of the whole organ and would be excellent especially
for teaching purposes. The inner structures could be studied in sub-organ level, with embedded serial sections. This is a non-trivial task since the structures would need to be connected throughout the sparse section stack in an automated manner.
With sections $50\mu m$ apart, a single structure has discontinuities between consecutive sections, making e.g. region growing algorithms inapplicable in the task. Yet, this is a natural next step of the VR prostate project.

The application can also be extended to cover other types of data besides the structure models and features. For example, we could present immunohistochemistry results in the application, embed individual nuclei to the models, or visualize spatial sequencing data of genomic or gene expression results acquired from the serial sections. We can also add more levels of detail to the application, for example new levels that submerge deeper into the tumor or even single cells. 

In this research, whole murine prostates were used as samples. However, other organs from other subjects, even human, could easily be visualized with the same methodology. The only requirement is that serial sections can be acquired and imaged
from the sample. With automated processing of the data, our method is easy to adopt for visualization of histological data from other sources.



\section*{Data acquisition and processing} 


\subsection*{Ethical permissions}

All animal experimentation and care procedures were carried out in accordance
with guidelines and regulations of the national Animal Experiment Board of Finland, and were approved by the board of laboratory animal work of the State Provincial Offices of South Finland (licence no \verb ESAVI/6271/04.10.03/2011 ) \cite{latonen2017vivo}.

\subsection*{Image acquisition}

Prostates of \emph{Pten+/-} mice \cite{cristofano1998pten} with intraglandular, noninvasive high grade PIN tumors were collected at 11 months, fixed in PAXgene$^{\textrm{TM}}$ (PreAnalytiX GmbH, Hombrechtikon, Switzerland) and embedded in paraffin \cite{latonen2017vivo}. Each prostate was cut into 5 $\mu m$ thick serial sections, which were placed on glass slides and HE-stained. The slides were scanned and stored in JPEG2000 format \cite{Tuominen2010}. The pixel size of the acquired WSIs was 0.46 $\mu m$.

Tumors were manually annotated with the freehand tool of ImageJ \cite{schneider2012nih}. Approximately 300 sections were acquired from a single prostate, from which every tenth section was HE-stained. Thus, each image stack has approximately 30 sections scanned into WSIs with 50 $\mu m$ distance between consecutive sections.

\subsection*{Image registration}

\begin{figure}
    \centering
    \includegraphics[width=\linewidth]{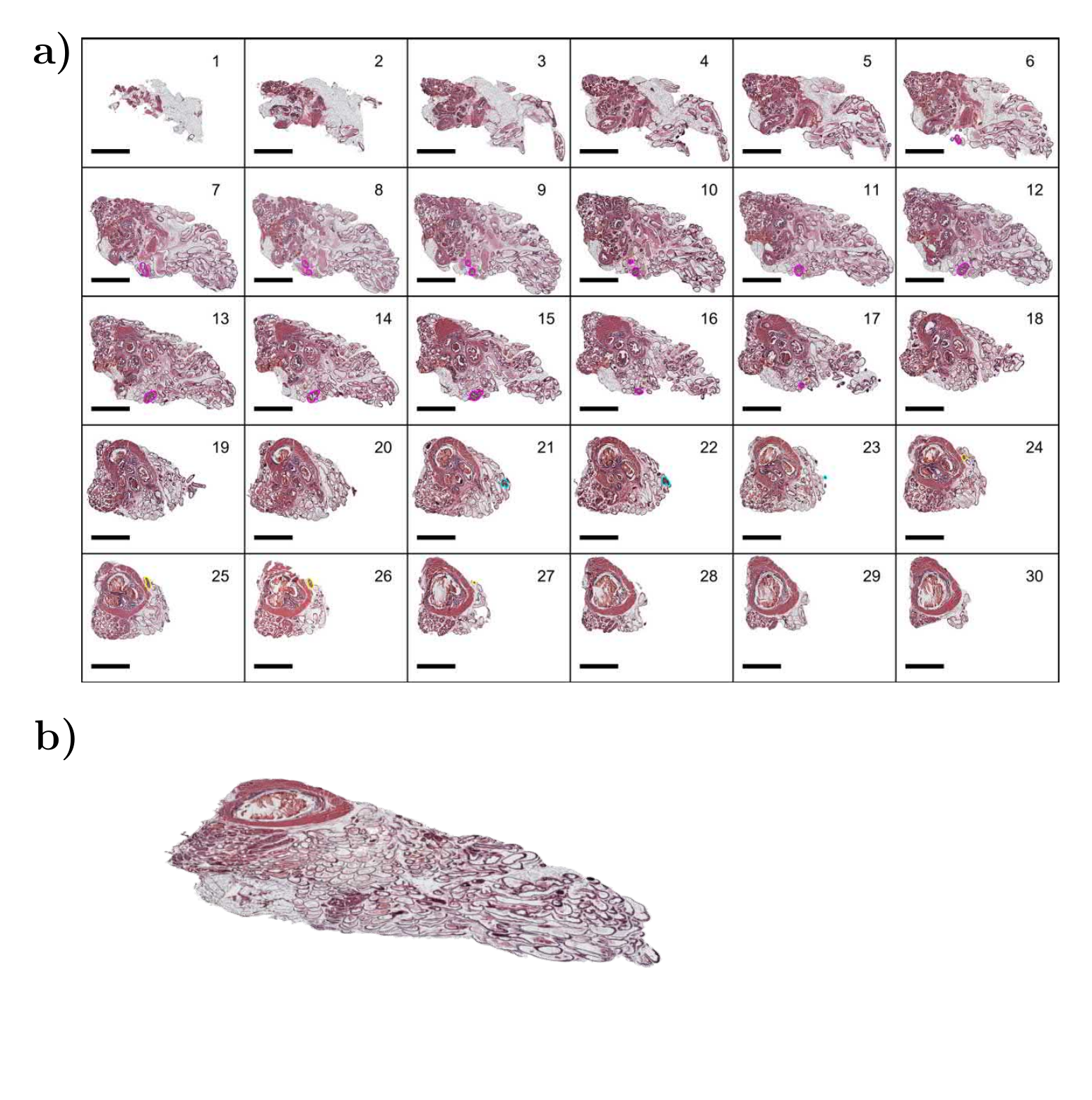}
    \caption{Every serial section of an example prostate illustrated, with tumors outlined in magenta, yellow and cyan (a) (scalebar = 1mm). Image stack visualized in 3D with Blender (b).}
    \label{fig:data}
\end{figure}

Histological sections have initially different relative orientations and locations in WSIs. Thus, an image registration step was performed to elastically co-register all images in a stack. The registration method searches corresponding locations in WSIs using normalized cross-correlation from adjacent sections and joins these together with virtual springs. Tissue's physical properties are also modeled with triangular spring mesh. This spring system is then allowed to bend the tissue until an equilibrium is found, denoting completed registration\cite{saalfeld2012elastic}. The final registration is obtained from the deformed spring mesh and a piece-wise linear transformation is applied accordingly. A Fiji implementation of this algorithm in TrakEm2 package was used \cite{cardona2012,schindelin2012fiji}. Resulting transformations were applied to binary masks of both tissue region and tumor annotations. All registered histological sections of one prostate sample with tumor annotations are presented in Figure~\ref{fig:data}(a), and the whole 3D stack after registration visualized in Figure~\ref{fig:data}(b).

\subsection*{Feature extraction}

The quantitative hand-crafted features, originally created for use with machine learning algorithms  \cite{ruusuvuori2016feature,valkonen2017analysis}, were computed with MATLAB R2017b \cite{MATLAB2017b}. The features include histograms of oriented gradients (HOG) \cite{dalal2005histograms}, local binary patterns (LBP) \cite{ojala2000gray}, amount of and distance between nuclei, and intensity features, to name a few. Features were computed from small image patches, with an approximate size of $400 \times 400$ pixels for features in the prostate level, and $100 \times 100$ for those in the tumor level.


\section*{Implementation}

\begin{figure}
    \centering
    \includegraphics[width=0.8\linewidth]{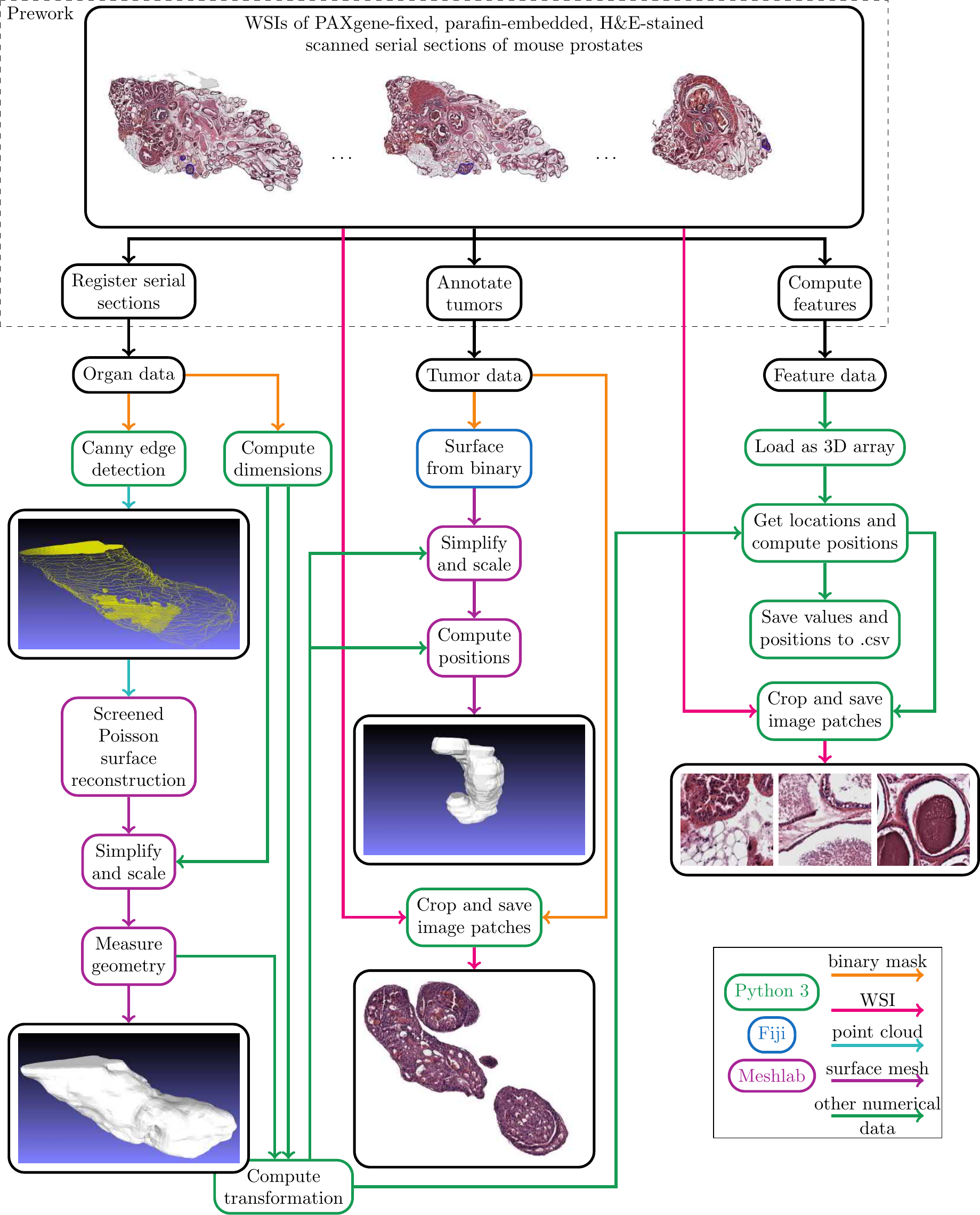}
    \caption{Data processing framework}
    \label{fig:framework}
\end{figure}

The implementation was automated as far as possible to enable visualization of new organ samples with minimal effort.
Data processing was mainly done with Python. Open source software FiJi (ImageJ)\cite{schindelin2012fiji} and Meshlab\cite{meshlab} were used for generating and processing the digital surface models (DSM). With data processing the original image stack data was transformed to formats that are fast and simple to load and process with a game engine. In addition, model and feature positioning was performed. Automated positioning of tumors and features in relation to the prostate is essential for combining the data in VR. The data processing steps are presented are presented in Figure~\ref{fig:framework}.
Unreal Engine 4 (version 4.21.2) \cite{ue4} (UE4) was used for VR development. Implementation was done with Blueprint Visual Scripting system and C++ programming.

\subsection*{Modeling of organs}


Organ DSMs were generated from binary masks with Screened Poisson surface reconstruction\cite{kazhdan2013screened} from 
point clouds using Meshlab. To be precise, MeshlabXML scripting interface \cite{meshlabxml} for Meshlab was used in Python, and Meshlab macros were recorded when functions not implemented in MeshlabXML were needed. 

Point cloud acquisition was performed in following manner. The full resolution binary masks were loaded and resized to 5\% of original size to reduce the amount of data points. Distance between serial sections in relation to image size was computed from known data properties (0.46$\mu$m pixel size and 50$\mu$m section distance). 
Then, borders of each mask were extracted with Canny edge detection and saved as a point cloud with correct vertical positions. To get planar surfaces from top and bottom layers, all points from these masks were included. 

For transforming the point cloud to a surface mesh, a Meshlab macro was recorded. First, the point cloud was loaded and point cloud simplification was performed. Then, normals for points were computed and smoothed. Finally, the point cloud was transformed into a mesh with Screened Poisson surface reconstruction and duplicate faces were removed. The macro was saved to be later called directly from Python. The reconstruction from point cloud results in smooth models with no trace of sparse, separate layers. 
While this method suits well for organ modeling, it fails with smaller objects like tumors. If object width is too small when compared to the distance between sections, the surface of the object is hard to define. Generated models might have ill-posed form and unwanted holes. 

The final processing of the model was performed with MeshlabXML. First, the model was scaled to fixed height. Since all samples have the same amount of sections, their size in VR is comparable if their height is the same. Then, rotation was performed to get suitable alignment. Finally, mesh properties of the organ model were acquired with \verb measure_geometry  function and saved to a properties file to be used with tumor models and features. 

\subsection*{Modeling of tumors}

Compared to whole organs, tumor volumes are typically fairly small, making them poorly compatible with similar Screened Poisson surface reconstruction that was used for whole organs. Thus, a Fiji macro was recorded to create models from tumor masks as follows. First, the masks were transformed to binary masks and dilated twice. The voxel depth was set to 1.2 from image properties to approximate correct tumor height. This step, however, is not essential since correct size will be computed later. 3D viewer was applied for creating the surface model, which was then saved as a binary STL file, as done previously in \cite{liimatainen20193d}. The data path was defined as an input parameter for the final macro file. Now, the Fiji macro was ready to be used from Python with \verb syscommand argument.


The properties of the prostate mesh were used to correctly scale and align the tumor models created with Fiji. The desired size $S_0$ and center position $P_0$ were computed from the relation between original tumor and prostate masks, and the size and position of the final prostate model. Tumor mesh center $P_1$ and size $S_1$ were acquired with \verb measure_geometry  function of MeshlabXML. The model was scaled with factor $S_0 / S_1$, and translated by $P_0 - P_1$. 

Finally, patches of full-sized serial sections were cropped to acquire histological images representing the tumor. The patches were acquired based on the furthest edges of tumor masks, using the tumor masks as alpha channels. Equally sized patches are easier to align in the VR application, and alpha channel is used to hide locations where tumor is not present. Images were saved as RGBA PNG format.

\subsection*{Feature processing}

The features used in this implementation were pre-computed with MATLAB as previously described \cite{ruusuvuori2016feature,valkonen2017analysis}. MATLAB feature arrays with location information were loaded to Python with scipy module's \verb loadmat  function. Features were loaded for each section separately, after which a 3D array was created to cover the whole prostate.

All necessary feature information was saved to comma-separated values (CSV) files for fast loading to Unreal Engine. For each pre-computed feature a separate CSV file was created. The location in relation to the prostate model was computed from scale and location information of the model. The information for each instance of each feature includes position, feature value, normalized feature value, and tumor information telling which tumor, if any, the feature resides in.

Each pixel in the feature map was given an individual index. The index is used to match the feature instance to the image patch from which it was computed from. The full-sized sections were cropped and patches were saved with corresponding indices as PNG files for fast image loading in UE4.


\subsection*{VR implementation for Oculus Rift with Unreal Engine 4}

\begin{figure}
    \centering
    \includegraphics[width=\linewidth]{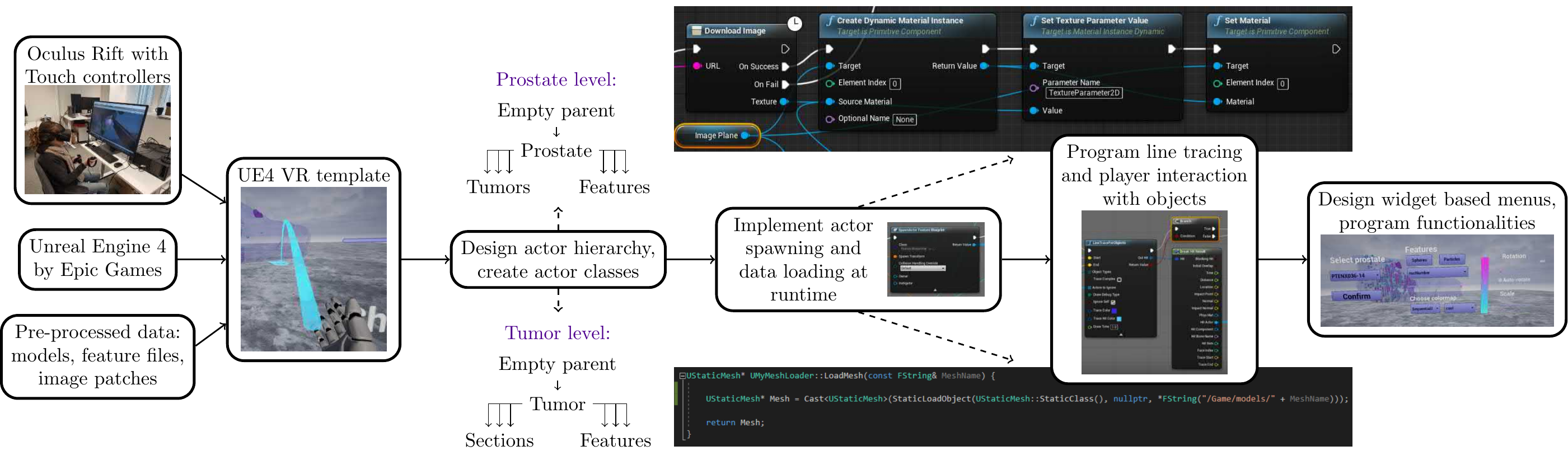}
    \caption{VR implementation steps.}
    \label{fig:vr_implementation}
\end{figure}

Most of the VR programming was performed with Blueprint Visual Scripting system in UE4. In addition, some necessary utilities were programmed with C++ and used as Blueprint nodes. VR implementation process is presented in Figure~\ref{fig:vr_implementation}.

Templates for character motion and controller mapping (\verb MotionControllerPawn,  \verb BP_MotionController)  were used as a starting point for VR implementation. These templates include teleportation movement, hand meshes and grabbing functions. Two levels were created, one for the whole organ and one for the detailed view of an object of interest. At this point, only basic level components like floor, walls, light, menu and player were added. All data, including models and features, are loaded dynamically during runtime. Organ and tumor models were imported to project Content file. After initial import, all models can easily be reimported if changes are made to models. Feature files and images were saved in an external folder from which they are loaded when needed. 

All data objects inherit the base Actor class of UE4. Intuitive Actor hierarchy was designed so that there is an empty parent for the main object in the level, which then parents Actors that lay inside the main object. The empty parent is spawned at the center of the main object based on mesh boundaries. All in-game positioning, scaling and rotation operations are applied to the empty parent, which all children in the hierarchy inherit while keeping their relative positions.
Actor-based Blueprint classes were created for each object type (prostate, tumor, feature). In addition, a game instance class was created for variables that need to be accessed from both levels. These variables, including prostate sample ID, tumor index and feature name, can be changed from an interactive widget menu in both levels.

Prostate and tumor Actors are spawned at level start. Since feature type and threshold-based visibility can be changed by the user, some optimization strategies were designed for the feature Actor class. Each feature has an ID corresponding to image patches, and this ID is the same for each different feature type at the same location. When a level (or a new prostate) is loaded for the first time, a map structure with feature ID as a key and reference to spawned feature Actor as a value is created. When the feature type is changed, we can simply query the map for the corresponding feature ID and change necessary properties (feature value, normalized value) of the feature Actor. When the threshold value is changed, we need to change visibility, collision properties, and color of the feature. However, changing these values for hundreds or thousands of features each time the slider moves makes the application lag. To create a smooth slider change where the user can see the effect in real-time, only the visibility of the feature is changed when the slider moves. Only when slider grab ends, collision properties and color of the feature are changed.

Interaction with objects is based on line tracing, where a direct line is drawn from hand mesh. If the line collides with an object, actions are performed. A laser beam from the hand is drawn for ease of use, otherwise it would be difficult to target e.g. a specific feature sphere. Actors to ignore -property of the line trace Blueprint node is used for excluding specific objects from line trace. For example, the right hand can grab spheres inside the prostate and tumors, and the left hand can only select tumors to switch to the tumor level. In the organ level, the user can grab features and the corresponding image patch is visualized. The image is loaded from the external folder and applied as a texture to the image plane Actor that is parented by the feature. When grabbing ends, the Actor with the image plane is destroyed to prevent memory problems resulting from too many textures.

Since this project concentrates on the visualization of scientific data, the colors play an important role. Thus, colormaps for feature visualization were provided and a colorbar was added to the menu. The user can select the preferred colormap from a widget-based dropdown menu. The colormaps were acquired from Python and a CSV file was loaded to Unreal Engine. The feature threshold can also be changed from the menu, and colors are scaled accordingly so that the full colormap is always in use. The prostate sample can be selected from the menu of the main level, and feature selection was implemented to both levels. In addition, changing an object's size and rotation was implemented, also to both levels.


\bibliography{vrrefs}



\section*{Acknowledgements}

We thank Hannu Hakkola for composing ambient music for the VR application.
We are grateful for Academy of Finland (P.R., grants \#313921 and \#314558) for funding the study. Also L.L. gratefully acknowledges funding from Academy of Finland (\#317871).

\section*{Author contributions statement}

K.L. designed and implemented the application. L.L. provided data annotations, histological knowledge, end-user testing and biological interpretation. M.V. and K.K. designed the pre-processing methods for segmentation and registration of images, and helped implement the feature extraction methods.
P.R. conceived and supervised the study, and implemented the feature extraction methods.
K.L., L.L. and P.R. wrote the manuscript. All authors reviewed and accepted the manuscript.







\end{document}